\documentclass[fleqn,12pt,twoside]{article}

\usepackage{graphicx}
\usepackage[figuresright]{rotating}

\newcommand{\AmS}{{\protect\the\textfont2
  A\kern-.1667em\lower.5ex\hbox{M}\kern-.125emS}}

\usepackage{espcrc1}

\title{First measurements of Collins and Sivers asymmetries at COMPASS}

\author{R. Webb\address[Erlangen]{Physikalisches Institut der Universit\"at
        Erlangen-N\"urnberg\\ Erwin-Rommel-Str 1, 91058 Erlangen, Germany} on
        behalf of the COMPASS collaboration}
       
\begin{document}

\maketitle

\begin{abstract}

COMPASS is a fixed-target experiment on the SPS M2 beamline at CERN. Its
$^{6}$LiD target can be polarised both longitudinally and transversally
with respect to the longitudinally polarised 160GeV/c $\mu^+$ beam. Approximately 20\% of the
beam-time in 2002, 2003 and 2004 was spent in the transverse configuration,
allowing the first measurement of both the Collins and Sivers asymmetries on a
deuterium target. First results from the the transverse data of the COMPASS
run in 2002 are reported here. 

\end{abstract}

\section{The Theoretical Background}

The cross-sections for polarised deep inelastic scattering \cite{Barone:2001sp}
of leptons on spin-1/2 hadrons can be expressed at leading twist as a
function of three independent quark distribution functions: the unpolarised
function $q (x)$, the longitudinally polarised ``helicity'' function $\Delta q
(x)$ and the transversely polarised or ``transversity'' function $\Delta_T q
(x)$. This latter function is chiral-odd and therefore decouples
from inclusive DIS. It can however be measured in semi-inclusive DIS where it
appears in combination with a chiral-odd fragmentation function, the Collins
function $\Delta D_a^h (z, p^h_T)$, in an azimuthal single-spin asymmetry
(SSA) in the hadronic end-product \cite{Collins:1992kk}. A similar asymmetry
could however also arise from a modulation of the transverse momentum $k_T$ of
unpolarised quarks in a transversely polarised nucleon represented by the
Sivers function, $\Delta^T_0 q$  \cite{Sivers:1990}. The Collins and Sivers effects can be
disentangled in leptoproduction on a transversely polarised nucleon, since
they exhibit a dependence on linearly independent kinematic variables.

The Collins hypothesis holds that the fragmentation function of a quark of
flavour $a$ in a hadron $h$ can be written as \cite{Baum:1996yv}:
\begin{equation}
D_a^h(z, {\bf p_T^{\, h}}) = D_a^h(z, p_T^{h}) + \Delta D_a^h(z, p_T^h) \cdot sin\Phi_C
\end{equation}
where ${\bf p_T^{\, h}}$ is the final hadron transverse momentum
with respect to the quark direction -- i.e. the virtual photon direction --
and $z = E_h / (E_{l}-E_{l'})$ is the fraction of available energy
carried by the hadron ($E_h$ is the hadron energy, and $E_{l}$ and $E_{l'}$
are the incoming and scattered lepton energies respectively).
The angle $\Phi_C$ appearing in the fragmentation function is known as
``Collins angle'' and is conveniently defined in the system where the
z-axis is the virtual photon direction and the x-z plane is the muon
scattering plane as $\Phi_C= \Phi_h -\Phi_s'$, where
$\Phi_h$ is the hadron azimuthal angle, and $\Phi_s'$ is the azimuthal angle
of the transverse spin of the struck quark (nucleon).
Since $\Phi_s'=\pi-\Phi_s$, where $\Phi_s$ is the azimuthal angle of the
transverse spin of the initial quark (nucleon), the relation $\Phi_C = \Phi_h
+\Phi_s - \pi$ is also valid. The fragmentation function $\Delta D_a^h(z,
p_T^{\, h})$ couples to transverse spin distribution function $\Delta_T q (x)$
and gives rise to an SSA (denoted as $A_{Coll}$) dependent on the kinematic
variables $x$, $z$ and $p_T^{\, h}$.

Following the Sivers hypothesis, the difference in the probability of finding
an unpolarised quark of transverse momentum $\bf k_T$ and $ - \bf k_T$ inside
a polarised nucleon can be written as \cite{Anselmino2002}:

\begin{equation}
{\mathrm P}_{q/p^\uparrow} (x, {\bf k_T}) - {\mathrm P}_{q/p^\uparrow} (x, - {\bf k_T}) =~
\sin\Phi_S~ \Delta^T_0 q ( x, k^2_T)
\end{equation}

where $\Phi_S= \Phi_k -\Phi_s$ is the azimuthal angle of the quark
with respect to the nucleon transverse spin orientation. It has been recently
demonstrated by theoretical arguments \cite{Brodsky:2002,Collins:2002}, that
the SSA (denoted by $A_{Siv}$) coming from the coupling of the Sivers function
with the unpolarised fragmentation function $D_a^h(z, p_T^{\, h})$ can be
observed at the leading twist from polarised semi-inclusive DIS.

\section{The COMPASS Experiment}

The COMPASS \cite{Baum:1996yv,Bressan:Spin2004} experiment on the M2 beamline
of the SPS accelerator at CERN makes use of a high
energy, intense $\mu^+$ beam naturally polarised by the pion-decay
mechanism. The helicity of the beam is negative, and the polarisation lies at
around 76\% at a momentum of 160 GeV/c. The spill structure of the SPS in 2002
consisted of a 4.8 s burst containing approximately $2 \cdot 10^8$ muons
followed by 11.4 s out-time. The luminosity of the beam was around $5 \cdot
10^{32}$ cm$^{-2}$s$^{-1}$. The spectrometer itself comprises two stages each
equipped with a spectrometer magnet and detectors for tracking, energy
determination and particle identification \cite{Mallot:2004}.

COMPASS uses the polarised target system of the SMC experiment
which consists of two $^6$LiD cells, each 60 cm long and with a radius of 1.5
cm located one after the other along the beam axis. In longitudinal running a
2.5 T field is maintained by a solenoid magnet, supplemented in transverse running by an orthogonal 0.5 T dipole field. The relaxation time of
around 2000 hours of the dipole field is sufficient to allow the transverse
measurement. The two target cells are always maintained with opposite
polarisation. Transverse polarisations $P$ of +/- 50 \% are routinely reached;
the dilution factor $f$ of the target material is calculated to be around 0.38.

\section{Analysis of the 2002 Data}

Two periods of data with transverse target polarisation were taken in 2002.
In order to compensate the different acceptances of the target cells, their
polarity was flipped by microwave reversal halfway through each period. The
counting rate asymmetry between these two sub-periods was then measured for
each target-cell separately. The number of events in each of the two
polarisation states ($\uparrow$/$\downarrow$) may be written as a function of
the Collins (Sivers) angle as  
\begin{equation}
N_{\uparrow \downarrow}(\Phi_{C/S}) = \alpha(\Phi_{C/S}) \cdot N_0 \, (1 \pm
\epsilon_{C/S} \sin{\Phi_{C/S}})\, ,
\end{equation}
\noindent where $\epsilon$ is the amplitude of the experimental asymmetry and
$\alpha$ is a function containing the acceptance of the apparatus. The angles
are always calculated assuming the target polarisation point up in the lab system.
The ``raw'' amplitude $\epsilon$ is connected to the Collins (Sivers) asymmetry through the expression
\begin{equation}
\epsilon_C = A_{Coll} \cdot P_T \cdot f \cdot D_{NN} \qquad or \qquad  \epsilon_S =
A_{Siv} \cdot P_T \cdot f\, ,\nonumber
\end{equation}
\noindent with $D_{NN}$ is the spin transfer coefficient or ``depolarisation
factor''. Indicating with $y$ the fractional energy
transfer from muon to virtual photon in the initial scattering process,
$D_{NN}$ may be calculated from the kinematics of each individual event through
$D_{NN} = (1-y)/(1-y-y^2/2)$ in the case of the Collins effect. For the Sivers
effect $D_{NN} = 1$, since this effect deals with unpolarised quarks.


In each period $\epsilon_{C}$ ($\epsilon_{S}$) is fitted
separately for the two target cells from the event flux with the two target
orientations using the expression
\begin{equation}
\epsilon_{C/S} \sin \Phi_{C/S} = \frac{N^{\uparrow}_h(\Phi_{C/S}) - R \cdot N^{\downarrow}_h(\Phi_{C/S})}
{N^{\uparrow}_h(\Phi_{C/S}) + R \cdot N^{\downarrow}_h(\Phi_{C/S})}
\end{equation}
where $R = N^{\uparrow}_{h, tot}/N^{\downarrow}_{h, tot}$ is the ratio of the
total number of events in the two target polarisation orientations.

The total sample collected in the 2002 transverse data-taking periods amounted
to about 200 pb$^{-1}$ in terms of integrated luminosity.  Events were
selected in which a primary vertex with identified beam and scattered muon and
at least one outgoing hadron 
was found in one of the two target cells. A
clean separation of muon and hadron samples was achieved by cuts on the amount
of material traversed in the spectrometer. In addition, the kinematic cuts
$Q^2 > 1$(G$e$V$/c)^2$, $W > 5$ G$e$V/$c^2$ and $0.1 < y < 0.9$ were
applied to the data to ensure a deep-inelastic sample above the region of
the nuclear resonances and within the COMPASS trigger acceptance. The upper
bound on $y$ also served to keep radiative corrections small. The asymmetries
were calculated for two different samples: both for all hadrons emerging from
the primary vertex, and, a sub-sample of this, only for the most energetic or
``leading'' hadron in each event. The leading hadron  was determined as the
most energetic non-muonic
particle of the primary vertex with an energy component $z > 0.25$ and a
transverse momentum $p_T^{\, h} > 0.1$ G$e$V/$c$. When all the hadrons coming
from the primary vertex were considered, the $z$ cut was lowered to
$0.20$. The final data sample had average values $ x = 0.034$, $ y  =
0.33$ and $ Q^2 =  2.7$ (G$e$V$/c)^2$. The average value for $z$ and $p_T^{\,
  h}$ are $0.44$ and $0.51$ G$e$V/$c$ respectively for the leading hadron analysis, and
$0.38$ and $0.48$ G$e$V$/c$ in the all hadrons case.

\begin{figure}[t] %
\begin{center}
\includegraphics[scale=0.45]{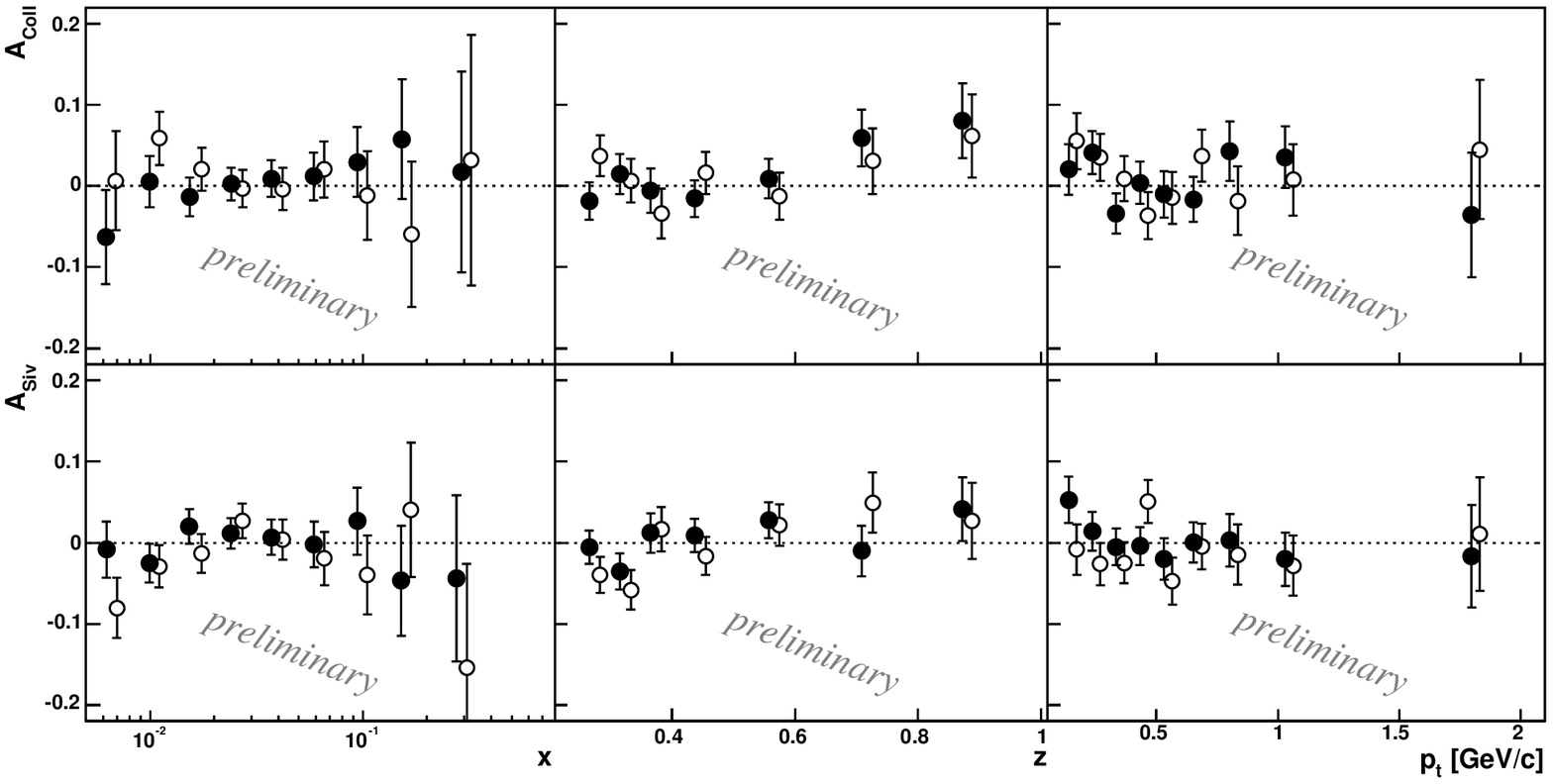}
\includegraphics[scale=0.45]{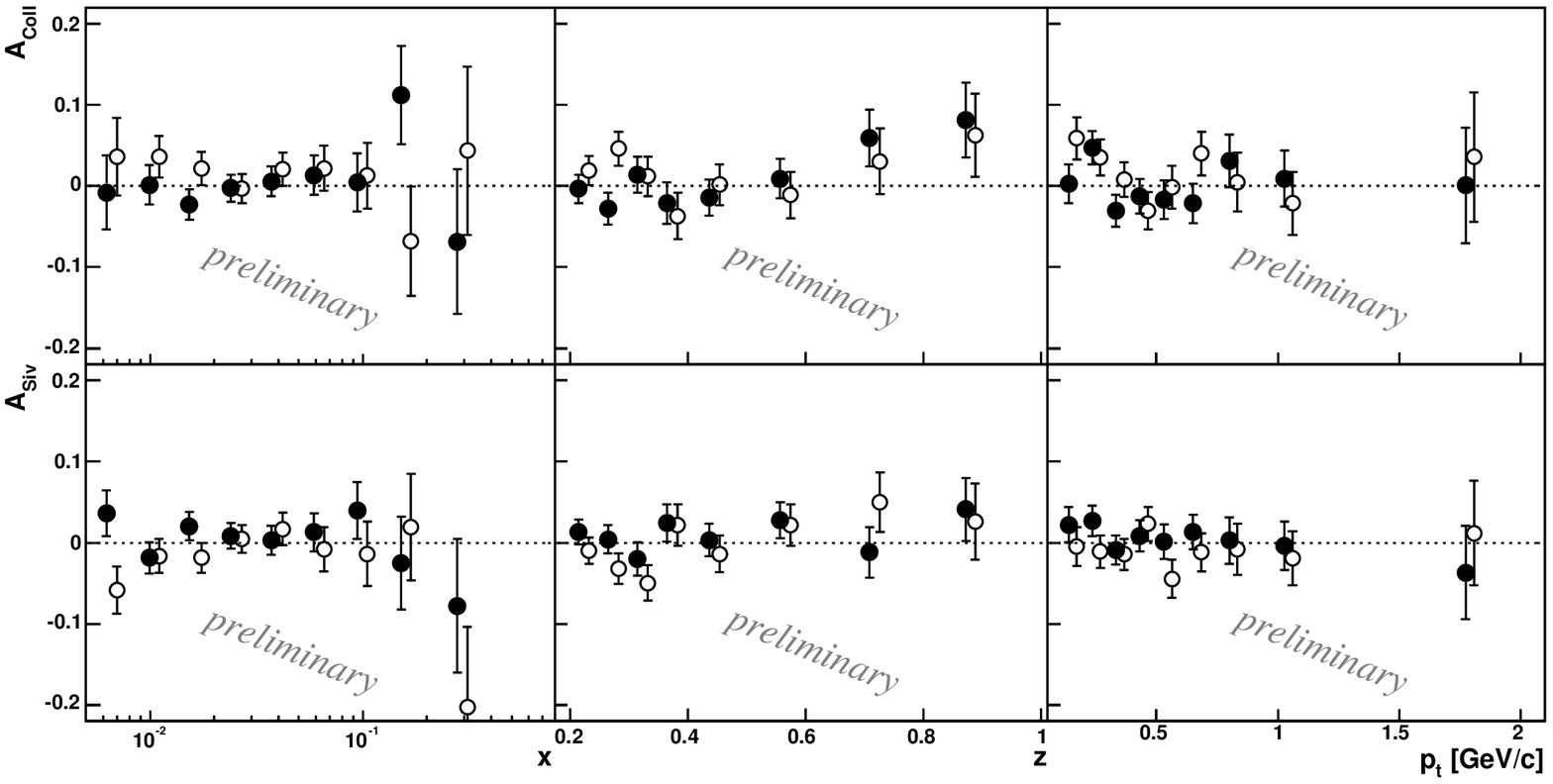}
\caption{Collins and Sivers asymmetry for positive (full points) and negative (open points) hadrons as a function of  $x$, $z$ and $p_T^{\, h}$. Leading hadrons analysis on top canvas, all hadrons on bottom canvas.}\label{results}
\end{center}
\end{figure}

The results of the asymmetries plotted against the kinematic variables $x$,
$z$ and $p_T^{\, h}$ are shown in Fig. \ref{results} for positive (full
points) and negative (open points) hadrons, and for the analysis of leading
(top plot) hadron and all hadrons (bottom plot).

Stability checks showed that the ratio of acceptances and efficiencies as a
function of the Collins (Sivers) angle does not change between two spin
orientations. Furthermore the results were stable when each target cell was
split into two parts and the asymmetry calculated separately, and when the
data was split into high and low hadron momenta. This leads to the conclusion
that systematic effects are smaller than the statistical errors.

In these first measurements of transverse spin effects on a deuteron
target, the Collins and
Sivers asymmetries were found to be compatible with zero within the statistical accuracy of the data. A
marginal indication of a Collins effect at large $z$ for both positive and
negative charges is seen. Coupled with HERMES results on a proton target also
reported at this conference, this may imply a negative neutron contribution to
the asymmetries.

A sensitivity improvement of factor two is expected once the data from the 2003
and 2004 COMPASS runs have been analysed. From 2006
onwards,  a target magnet with a significantly larger acceptance should
increase the statistics at high $x$. Complementary measurements with a
proton target are also planned.


\begin{thebibliography}{9}
\bibitem{Barone:2001sp}
V. Barone, A. Drago and P. G. Ratcliffe,
Phys.\ Rept.\  {\bf 359} (2002) 
\bibitem{Collins:1992kk}
J. C. Collins,
Nucl.\ Phys.\ B {\bf 396} (1993) 161.
\bibitem{Sivers:1990}
D. W. Sivers,
Phys.\ Rev.\ D {\bf 41} (1990) 83.
\bibitem{Baum:1996yv}
G. Baum {\it et al.}  [COMPASS Collaboration],
CERN-SPSLC-96-14.
\bibitem{Anselmino2002}
M. Anselmino,  V. Barone, A. Drago,  F. Murgia,
[arXiv:hep-ph/0209073].
\bibitem{Brodsky:2002}
S. J. Brodsky, D.S. Hwang, I. Schmidt,
Phys.\ Lett.\  B {\bf 530} (2002) 99.
\bibitem{Collins:2002}
J. C. Collins,
Phys.\ Lett.\  B {\bf 536} (2002) 43.
\bibitem{Bressan:Spin2004}
Y Bedfer {\it et al.} , ``Recent COMPASS spin physics results'', these proceedings.
\bibitem{Mallot:2004}
G.K. Mallot, NIM {\bf 518}(2004) 121.
\end{thebibliography}
\end{document}